\documentclass[letterpaper,twocolumn,10pt]{article}
\usepackage{graphics}
\usepackage{myusenix}
\usepackage{url}
\usepackage{verbatim}
\usepackage{algorithm}
\usepackage[noend]{algorithmic}
\usepackage{fitch}
\usepackage{subfigure}

\author{Petros Maniatis \hspace{4mm} Mary Baker\\ {\em Computer
Science Department, Stanford University}\\ {\em Stanford, CA 94305,
USA}\\
{\tt \{maniatis,mgbaker\}@cs.stanford.edu}\\
\url{http://identiscape.stanford.edu/}} 

\date{}

\newcommand{\key}[1]{{\tt#1\ldots}}
\newcommand{\pn}[2]{\textsf{Primary}[\url{#1}, \url{#2}]}
\renewcommand{\pn}[2]{\url{#1@#2}}
\newcommand{\ihn}[3]{\textsf{Historic}[\url{#1}, \url{#2}, #3]}
\renewcommand{\ihn}[3]{\textsf{Historic}[\pn{#1}{#2}, #3]}
\newcommand{\ha}[1]{$\mathcal{#1}$}
\newcommand{\now}{\ensuremath{\mathit{now}}}

\newcommand{\pof}[1]{\ensuremath{\mathit{Person}(#1)}}
\newcommand{\nof}[1]{\ensuremath{\mathit{Name}(#1)}}

\newcommand{\stimeof}[1]{\ensuremath{\mathit{StartTime}(#1)}}

\newcommand{\extimeof}[1]{\ensuremath{\mathit{ExpirationTime}(#1)}}

\begin{document}

\title{\bf A Historic Name-Trail Service}

\maketitle

\thispagestyle{empty}

\begin{abstract}

\small 

People change the identifiers through which they are reachable online as
they change jobs or residences or Internet service providers.  This kind
of personal mobility makes reaching people online error-prone.  As
people move, they do not always know who or what has cached their now
obsolete identifiers so as to inform them of the move.  Use of these old
identifiers can cause delivery failure of important messages, or worse,
may cause delivery of messages to unintended recipients.  For example, a
sensitive email message sent to my now obsolete work address at a former
place of employment may reach my unfriendly former boss instead of me.

In this paper we describe HINTS, a historic name-trail service.  This
service provides a persistent way to name willing participants online
using today's transient online identifiers.  HINTS accomplishes this by
connecting together the names a person uses along with the times during
which those names were valid for the person, thus giving people control
over the historic use of their names.  A correspondent who wishes to
reach a mobile person can use an obsolete online name for that person,
qualified with a time at which the online name was successfully used;
HINTS resolves this historic name to a current valid online identifier
for the intended recipient, if that recipient has chosen to leave a name
trail in HINTS.


\end{abstract}

\section{\label{sec:Introduction}Introduction}

The online world is inhabited by nomads.  People change their online
names as they switch Internet service providers (ISPs), either because they
change jobs and they use the ISP of their employer, or because they
switch to a better, cheaper, or more convenient service for their
personal ISP.  As a result, people accumulate a legacy of online
identifiers that are, in most cases, dangling pointers into
obscurity, making it hard to reach people.

Unfortunately, sometimes an obsolete online identifier does point
to something, and this can be much more dangerous than a dangling
identifier.  If the obsolete identifier belongs to my previous
personal ISP, then some unrelated, unfortunate subscriber of that
same ISP is bothered by my legacy of spam.  If the obsolete
identifier belongs to my previous place of employment, it might
allow my sensitive communications to reach a potentially disgruntled
former colleague or boss.

The problem stems from the simple fact that online identifiers \emph{do
not} belong the people that they name; they belong to the organization
that manages the associated name space.  An employee of Sample
University does not own her Sample U.\ email address; that address
belongs to the university itself, which loans it out to its employee for
the duration of her employment there.  Similarly, my \url{yahoo.com}
address only names me as long as Yahoo! allows me to use it and
maintains its service.  This is essentially a mobility problem: a mobile
person moves slowly from identifier to identifier in a
potentially-changing landscape of online names over which he has little
control or authority.  We call this the \emph{personal-identity
mobility} problem.

The promise of unique personal identifiers as a panacea for all kinds of
identity mobility has been made many times with dubious success.  In
some cases, the proposed identification scheme requires a retrofit of
the entire communications infrastructure of the Internet to
work~\cite{Maniatis2000}.  In other cases, the ``unique personal
identifier'' is just another identifier assigned from a proprietary name
space~\cite{CommonName,DigitalMe,OneName,Pingid}.

In this paper we propose the {\bf HI}storic {\bf N}ame-{\bf T}rail {\bf
S}ervice (HINTS).  In our naming scheme, current online identifiers,
email addresses and instant messaging account names remain the primary
identifiers that people use in everyday communications.  However,
HINTS enables a person voluntarily to build a trail that connects all the
identifiers that person uses over time into a single \emph{name history}.
A correspondent can then safely name the person by using a potentially obsolete
identifier qualified with a time at which that identifier was
successfully used in the past.  With this information the appropriate
name history for a unique person can be found and followed to its latest
entries, which provide a more recent way to reach the sought person.

HINTS offers mobile people control over forward pointers from their old
to their current online identifiers.  Even if an old name
belongs in a name space maintained by a now-defunct organization,
that name can still be resolved within HINTS to a current identifier
for its past owner.  This independence of a historical name from the
provider of its associated name space promotes a robust, persistent
way to refer to people, even when the practical names we use in
everyday life are neither robust, unique nor persistent.

The goal of this paper is to introduce the concept of historic naming as
a means to address identity mobility, and to describe a simple version
of HINTS that can be deployed using today's technology.  However, our
current efforts focus on designing and implementing a decentralized,
more fault tolerant and secure version of HINTS, which we also briefly
describe in this paper.

We proceed by providing an abstract model of the current personal online
identification landscape, in Section~\ref{sec:model}.  In
Section~\ref{sec:historicNameService} we introduce the basic concepts of
HINTS, including the structure of the name space, the functionality of
the service, and the implementation considerations involved.  We
evaluate the benefits offered by HINTS in Section~\ref{sec:discussion},
by delineating what the service can and cannot do, given how personal
online identification works today.  We then elaborate on the next step
for HINTS, namely security and fault-tolerance, we discuss what
additional requirements these properties place on online service
providers, and we sketch an enhanced design in
Section~\ref{sec:secureHints}, before we conclude.

\section{A Model For Personal Online Identification}
\label{sec:model}

This section outlines the naming model on which the historic
name-trail service is based.  This model abstracts the way in which
names are \emph{currently} associated with people online.  We
also describe the components that currently provide naming services
and explain the division of control and naming responsibilities between
these services and their clients.

People need names online for a variety of purposes, such as
authentication, communication, and authorization.  In most cases, these
names are simple, such as plain email addresses~\cite{RFC0821}.  In
other cases, online names may be two-tiered, starting with a primary
application-unspecific name, which is used to look up
application-specific addresses in a directory via a directory access
protocol such as LDAP~\cite{RFC2251}.  For simplicity, we take the
former approach in this paper.  Specifically, we use email addresses as
the \emph{primary} names of people online in our examples.  However, the
techniques we describe in later sections apply directly and without
change to names in other applications, such as instant messaging, or to
two-tiered naming environments.

The names people use are drawn from a multitude of independent
\emph{name spaces}.  A name space is a set of all possible names that
are or can be assigned by a single administrative entity, called a
\emph{name-space provider}.  For example, Yahoo!\ Inc.\ is the
name-space provider operating the \url{yahoo.com} name space: this is
the set of all possible account names for Yahoo!\ Inc.'s services, such
as web mail, instant messaging, calendaring, etc.  In most cases, the
name-space provider responsible for a name is easy to determine from the
name itself.  For example, \pn{name}{yahoo.com} is a name in Yahoo!'s
name space.  The computers responsible for maintaining a name space are
the \emph{name servers} of the provider.  To find an authoritative name
server for a name space, we can access the Domain Name
Service~\cite{RFC1034}, either through a distinct DNS resource record
type, or through a straightforward name mapping (e.g.,
\url{names.sample.edu} for the \url{sample.edu} name space).

A person online may be addressable via multiple names drawn from
different name spaces.  For example, one may be addressed using
names assigned by a school, by a personal Internet service provider,
by web services such as Yahoo!\ and Hotmail, and by professional
associations such as the ACM.  These names can be used for distinct
or overlapping purposes: professional, personal, commercial.

The association of a name with a person follows the regulations defined
by the corresponding name-space provider.  These regulations may dictate
whether a name association is temporary or permanent, whether a name may
be reassigned after its previous association is discontinued and, if so,
the minimum amount of time between successive assignments of a single
name.  For example, Sample University's Computer Science Department
assigns names to its graduates for life, whereas ISPs assign names to
clients for the duration of the pertinent service agreement.  It is
important to note, however, that the name-space provider is the ultimate
controller of the name space, regardless of the ``promises'' it makes to
those using its names.  It is not uncommon, for example, that a claimed
permanent name assignment has had to change for legal, social or
political reasons.  As an example experienced by one of the authors at
Sample University, the arrival of a new faculty member who desired an
email address already assigned to a graduate resulted in a reassignment
of that email address to the faculty member, in spite of the
institution's guarantee of lifelong identifier assignments to its
graduates.

In most cases today, a person exercises implicitly his ``stewardship''
of a name assigned to him, by being able to access the application for
which the name is assigned.  For example, in today's Internet, if I can
read and send email as \pn{name}{yahoo.com}, then I am assumed to be the
person to whom that name has been assigned.  Though not bulletproof,
this simplistic form of authentication is widely used for signing up in
mailing lists, for signing up for web services or even for conducting
business online.  The name-space provider can revoke an association
between a person and his name by stopping that person from accessing the
named service; for example, the provider can cancel the associated web
email account, or change the requisite password.  Aside from
intra-enterprise settings, not many name-space providers today offer
rigorous security in how they assign names, and how their clients assert
ownership on their names.  Section~\ref{sec:secureHints} describes a
more secure naming model, very similar to an attribute certification
model, that allows us to address the problem more comprehensively.

In the remainder of this paper, we limit \emph{mobile person} to be
someone who wishes to remain reachable despite identifier changes.  A
\emph{correspondent person} is someone who wishes to reach a mobile
person. In the examples we use, Jane Mobile is a mobile person, and Dan
Friend is a correspondent of Jane's.

\section{A Historic Name-Trail Service}
\label{sec:historicNameService}

The primary goal of this work is to provide mobile people with a
forwarding service that is available to help them remain reachable
despite their identity mobility.  We accomplish this by allowing a
mobile person to determine to what his historic names---names qualified
with a time when they validly named that person---point.  We approach
the problem first by taking the simplest path possible: a centralized,
trusted service that operates similarly to most web services currently
deployed and works with no cooperation from current infrastructure.  We
explore more sophisticated, secure and fault-tolerant possibilities
in Section~\ref{sec:secureHints}.

Reachability despite identity mobility requires the transfer of some of
the naming ``power'' over a name from the associated name-space provider
to the person to whom the provider assigns the name, within closely
guarded \emph{temporal} confines.  Specifically, even though a provider
can do anything it wishes to a name, its actions cannot be applied
retroactively: if Yahoo!  assigns \pn{jmobile}{yahoo.com} to Jane Mobile
from August 1999 to May 2000, it cannot later ``change history'' so as
to strip Jane of her control over \pn{jmobile}{yahoo.com} for the
indicated time period; Jane's authority over \pn{jmobile}{yahoo.com}
from August 1999 to May 2000 is \emph{persistent}.

By imposing this persistence of authority, HINTS splits
responsibilities between mobile people and name-space providers.
Name-space providers are responsible for creating and destroying
associations between names and people (in fact, between names and people
who can access the service state for that name).  People are responsible
for acting on behalf of, and being reachable as a particular name during
the period that they have been assigned that name.
 
This separation of control enables the definition of a ``virtual''
global persistent name space, with most of the ambitious properties
suggested in IdentiScape~\cite{Maniatis2000}, such as persistence,
controllability and human-centricity, but without requiring the creation
and maintenance of a centralized name service for a global, flat name
space implied by that system.

\subsection{The Name Space}
\label{sec:nameSpace}

We define the HINTS name space by extending names with a continuous time
designation.  For example, to refer to the person named by the
identifier \pn{jmobile}{yahoo.com} in March of 2000 we construct the
identifier \ihn{jmobile}{yahoo.com}{03/2000}.  More generally, the HINTS
name space contains identifiers of the form \ihn{name}{name
space}{time}.  A HINTS name corresponds to a time-specific primary name,
and is meaningful both while the associated primary name is valid (i.e.,
assigned), and after that primary name has been reassigned or obsoleted
by its name-space provider.

The time component of the identifier defines a version of the chosen
name space at a particular, coarse-grained time.  The time component may
designate an entire year, a year and a month, or a full date.  Since
identifiers change at ``human'' time scales, refining the time component
to anything shorter than a day may be unnecessary.  We assume that
clocks of clients or providers are at least coarsely synchronized (e.g.,
they all agree on what day it is), but we address a more secure timing
environment in Section~\ref{sec:secureHints}.

Many different HINTS names may denote the same person.  Since Jane
Mobile held the name \pn{jmobile}{yahoo.com} from August 1999 to May
2000, any HINTS name that corresponds to the same primary name and with
a time designation within the validity period points to her.  For
example, \ihn{jmobile}{yahoo.com}{9/8/1999} and
\ihn{jmobile}{yahoo.com}{1/2000} are valid HINTS names for Jane Mobile.

Some HINTS names denote multiple people.  Consider the scenario in which
Yahoo!\ Inc.\ reassigned the name \pn{jmobile}{yahoo.com} to a different
person, James Mobilewski, in September of 2000.  Then the HINTS name
\ihn{jmobile}{yahoo.com}{2000} is \emph{multivalent}, in the sense that
it points to two different people, Jane Mobile and James Mobilewski.  A
multivalent HINTS name is essentially a list name, since it names
\emph{all} of the possible people to whom the included primary name
points during the time designation of the historic name.  Therefore,
historic names with narrow time designations are preferable to those
with very broad time designations, when possible.

\subsection{A Personal Naming History}
\label{sec:nameHistory}

\begin{figure}
\centerline{\includegraphics{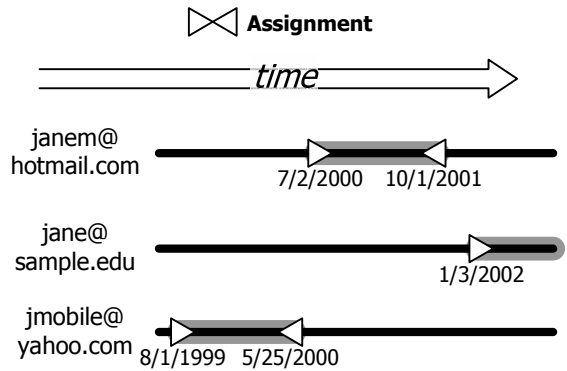}}
\caption[Jane's name history]{\small Jane's name history.  Each of the
three names shown has been held by Jane at some point in time.  For each
name, the thick gray line represents the time period during which Jane
was assigned that name by Yahoo!.  For example, Jane held
\pn{janem}{hotmail.com} between July 2000 and October 2001.}
\label{fig:NameTimeGraph}
\end{figure}

HINTS relies on a \emph{name history} to resolve historic names.  A name
history links together HINTS names that refer to the same person.  The
objective behind maintaining a name history is to be able to reach a
currently valid name-to-person association, starting with a now obsolete
one.  In the Jane Mobile example, the goal is to be able to obtain name
\pn{janem}{hotmail.com}, which is currently valid and is held by Jane,
starting with \ihn{jmobile}{yahoo.com}{03/2002}, even though
\pn{jmobile}{yahoo.com} is no longer held by Jane.
Figure~\ref{fig:NameTimeGraph} shows an example name history for Jane
Mobile.

A name history is maintained as a sequence of historic records
pertaining to a single name within a single name space, such as
\pn{jmobile}{yahoo.com}.  An online entity called a \emph{name
historian} maintains these records.  The history of a name is modified
due to either assignment changes or linking changes.

Changes in assignment affect to whom a particular name points.  These
are effected in provider-specific ways, such as changes in the password
needed to receive service under a particular name, etc.  Assignment
changes can be detected by the name historian either directly, through
an explicit notification from the name-space provider, or indirectly,
through failure of a mobile person to respond to a challenge issued to
his formerly assigned name.  Since in this design we assume no
cooperation from name-space providers, only indirect detection of
assignment changes is possible.  However, the historian can initiate
indirect detections of assignment changes reactively, at the request of
a mobile person.

Changes in linking affect how a single person assumes different online
names over time.  The historian emulates the concept of a ``person''
online by maintaining a trail of ``personal manifestations'', such as
knowledge of the same secret password or of the secret portion of the
same cryptographic key pair.  Linking represents the intention of such
an online person to be or not to be named by particular online names.
The mobile person requests explicitly to be linked to or unlinked from a
name by contacting the name historian with the appropriate request.

Assignments and links must both support the association of an online
name with the historian's online representation of a mobile person.
Specifically, to accept such an association, the historian must
establish two facts: first that the mobile person wishes to assume that
name, as communicated directly to the historian via a linking request;
\emph{and}, second, that the online person has access to that name,
which the historian can detect indirectly by sending a challenge to that
name via email (or any other applicable protocol) and expecting an
appropriate response.

The historian periodically reestablishes both assignment and linking for
an association.  Since the name-space providers are not aware of the
service, the historian has no recourse but to poll the name space
periodically or in response to user requests, so as to establish whether
the assumed mobile person still has control of his former name.
Similarly, to avoid cases where a mobile person who becomes unavailable
is assumed to be asserting control over a name by default for extended
time periods, the historian expects periodically that person to reassert
his links actively.

Figure~\ref{fig:AssociationRecord} illustrates an \emph{association
record}, the basic building block of a historic name database.  The
specific record identifies Jane Mobile using her first and last names,
although in practice a person is identified by the historian using an
internal, private, implementation-specific name space for mobile person
identification that no correspondent ever sees.  An association record
is active if it covers the present time (that is, its expiration time is
in the future).  An expired association record is archived and becomes
immutable.

\begin{figure}
\begin{verbatim}
AssociationRecord{
  Name            : jmobile@yahoo.com
  Person          : Jane Mobile
  Start Time      : March 2, 2000
  End Time        : May 1, 2000
  Expiration Time : July 1, 2000
  Next Link       : June 29, 2000
  Next Assign     : June 29, 2000
}
\end{verbatim}
\caption[An association record]{\small An association record
representing the association between Jane Mobile and the name
\pn{jmobile}{yahoo.com} from March 2 to May 1 of 2000.  The historian
considers the association valid until July 1 and expects a
reconfirmation thereof on June 29.  If that confirmation does not
arrive, the association record is archived as ending on May 1st.
Otherwise, the association record remains active, pushing its end time
to July 1st and its expiration time 2 months later.  The duration of the
time-to-live period of 2 months is arbitrarily chosen here and can be
modified per name space, per person, or per name historian.}
\label{fig:AssociationRecord}
\end{figure}

\begin{figure}
\centerline{\includegraphics{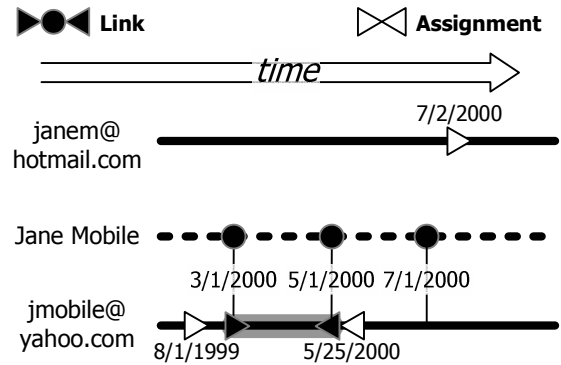}}
\caption[A portion of Jane Mobile's name history]{\small The names
\pn{jmobile}{yahoo.com} and \pn{janem}{hotmail.com} are linked to the
historian's account for Jane Mobile (dashed line in the middle).  The
gray line segment covers associations that the historian has accepted.}
\label{fig:Associations}
\end{figure}

Figure~\ref{fig:Associations} illustrates some of the historic
associations that the name historian maintains for Jane.  As shown in
Figure~\ref{fig:NameTimeGraph}, the name-space provider for
\pn{jmobile}{yahoo.com} unassigns the name from Jane on May 25, 2000.
As a result, even though Jane wishes to retain the name, as evidenced by
the link extension she requested to July 1st, the historian stores an
archived association record that only extends up to the last time both
the name-space provider and Jane agreed on the association, May 1, 2000.

\subsection{The Name Historian}
\label{sec:historian}

In this section, we delineate the functionality available to users of
HINTS.  In particular, we describe the interface to the name historian
itself, its design and some implementation details.

The primary objective of any name service is to support name resolution:
given a personal identifier, the service must return a ``pointer'' to
the identified person.  In the HINTS context, this means that HINTS
names must be resolved to primary names, such as email addresses.  If
Dan Sender wishes to send email to Jane Mobile, he must first resolve
the HINTS name with which he previously reached her successfully
(\ihn{jmobile}{yahoo.com}{03/2000}) to a currently valid primary name
(\pn{jane}{sample.edu}).  Keeping track of the last time an email
address was used successfully is arguably functionality that current
email clients can easily offer, and is very similar in complexity to
keeping track, for example, of the last time a web URL was used, which
most web browsers do transparently by default.  Alternatively,
correspondents can remember an approximate time period when they
contacted a mobile person and construct a historic name accordingly,
perhaps by referring to saved correspondence with the mobile person and
extracting dates from that correspondence.

The name historian runs a centralized, trusted service maintaining name
histories.  The historian is trusted to check the validity of the name
histories it stores and to report on those histories when asked.  Given
the naming model we assume in this design (see Section~\ref{sec:model}),
the historian is a powerful component in the network.  In
Section~\ref{sec:discussion} we examine the shortcomings of having such
a centralized design, and we propose alternatives in
Section~\ref{sec:secureHints}, based on stronger assumptions about
name-space providers.

The historian operates a name space itself, consisting of the account
names of its users.  This name space, however, is only used to
authenticate mobile people to the historian and need not be visible
externally to the service.  The historian manages its own name space
similarly to how most web services manage their account name spaces: a
client (mobile person in this case) signs up online and receives an
account and a means of authentication, such as a password or an
asymmetric key pair, for future exchanges with the historian.

To accomplish the task of resolution, the historian must also be able to
receive link requests from mobile people as described in the previous
section, and send and verify challenges to the holders of primary names
to detect name assignment changes.

HINTS can help correspondents who must rely on their own memory to
construct a resolution request by presenting them with a history of name
associations for the same name.  For example, Dan Friend remembers
having contacted Jane in the late 90's, but that is as specific a time
frame as he can recall.  If he can find out that \pn{jmobile}{yahoo.com}
belonged to someone from 1995 to 2000, and then to someone else from
2000 to 2002, then Dan can reasonably choose
\ihn{jmobile}{yahoo.com}{1999} to locate Jane Mobile.

In summary, the history service must be able to perform the following
tasks:
\begin{itemize}
\item Create mobile person accounts.
\item Request link changes (linking and unlinking) between a mobile
person account and a primary name.
\item Resolve a historic name to a currently assigned primary name.
\item Retrieve a list of association periods for a primary name.
\end{itemize}

The historian maintains a database of association records, as described
in the previous section.  The database maintains two sorted indices, one
on [\emph{name}, \emph{start time}] and another on [\emph{person},
\emph{start time}].

When a mobile person requests to be linked to a name, the historian
contacts that name with a random number as a challenge, by email or any
other applicable protocol.  If the email is returned with the
appropriate response to the challenge, then the historian considers the
name assigned to the mobile person and creates a new association record.
Algorithm~\ref{alg:createLink} details this process.

\begin{algorithm}
\caption[Linking a primary name to a mobile person]{Link mobile person
\ha{A} to name \pn{a}{A}. $H$ is the historian, $M$ is the mobile
person, $\now$ is the current time (on the historian's clock) and $l$ is
the maximum validity period of a link.  $M$ is assumed to have logged in
as \ha{A}.  $\Rightarrow$ denotes a network message.  $\leftarrow$
denotes assignment.  The process executes on $H$.}
\label{alg:createLink}
\begin{algorithmic}[1]
\sffamily

\STATE $H \Leftarrow M$: Link \ha{A} to \pn{a}{A}

\STATE $N \leftarrow$ random nonce

\STATE $H \Rightarrow $\pn{a}{A}: Request confirmation of assignment to
\ha{A} with nonce $N$

\STATE $H \Leftarrow M$: Assignment confirmation of \pn{a}{A} to \ha{A}
with nonce $N$

\IF{confirmation is negative or none arrives}

\STATE Do nothing and exit

\ENDIF

\STATE $L \leftarrow$ the association records where name is \pn{a}{A}
and person is \ha{A} with the greatest start time

\IF{none is found}

\STATE Store into database [Association \pn{a}{A}, \ha{A}, $\now$,
$\now$, $\now$ + $l$, $\now$ + $l$, $\now$ + $l$] \COMMENT{format is
name, person, start, end, expiration, next link, next assignment}

\ELSE

\IF{$\extimeof{L} < \now$}

\STATE Store into database [Association \pn{a}{A}, \ha{A}, $\now$,
$\now$, $\now$ + $l$, $\now$ + $l$, $\now$ + $l$]

\ELSE

\STATE Update into database record $L$ to [Association \pn{a}{A},
\ha{A}, \stimeof{L}, $\now$, $\now$ + $l$, $\now$ + $l$, $\now$ + $l$]

\ENDIF
\ENDIF

\end{algorithmic}
\end{algorithm}

Links are severed in a similar manner, although no confirmation from the
associated primary name is necessary; the reason for severing a link is
exactly that the associated primary name is no longer under the control
of the mobile person and, as a result, a confirmation from the primary
name might not even be possible.  However, a notification is sent to the
primary name, to make it harder for an unauthorized user of the
historian account to make changes unobtrusively.
Algorithm~\ref{alg:simpleSeverance} describes the process in more
detail.

\begin{algorithm}
\caption[Severing a primary name from a historian account in the simple
historic naming scheme]{Sever the link from the historian account \ha{A}
to name \pn{a}{A}. $H$ is the historian and $M$ is the mobile person,
logged in as \ha{A}.}
\label{alg:simpleSeverance}
\begin{algorithmic}[1]
\sffamily

\STATE $H \Leftarrow M$: Sever \pn{a}{A} from \ha{A}

\STATE $H \Rightarrow$ \pn{a}{A}: Notification of link severance from
\ha{A}

\STATE $L \leftarrow$ the association record where name is \pn{a}{A},
person is \ha{A} with the greatest start time

\IF{$\extimeof{L}>\now$}

\STATE Update into database record $L$ to [Association \pn{a}{A},
\ha{A}, \stimeof{L}, \now, \now, \now, \now]

\ENDIF

\end{algorithmic}
\end{algorithm}

Correspondents query the historian for resolved historic names in a
simple request-response protocol.  The historian resolves a historic
name by, first mapping it to an internal account and then finding the
latest still-valid association from that account to a primary name.
Algorithm~\ref{alg:simpleResolution} has the details.

\begin{algorithm}
\caption[Name resolution from a historic name to a primary name]{Resolve
historic name \ihn{a}{A}{t} to primary name \pn{b}{B} if possible, or
return failure otherwise.  This is run on the historian $H$.}
\label{alg:simpleResolution}
\begin{algorithmic}[1]
\sffamily

\STATE $L\leftarrow$ the association record where name is \pn{a}{A} and
the start time $s$ is the highest such that $s\leq t$

\IF{no record is found}

\STATE Return no results

\ENDIF

\STATE $L'\leftarrow$ the association record where person is \pof{L} and
the end time is greater than \now

\IF{no record is found}

\STATE Return no results

\ENDIF

\STATE Return \nof{L'}

\end{algorithmic}
\end{algorithm}

Listing association periods for a name is a straightforward operation
helped by the database indices maintained.

\section{Discussion}
\label{sec:discussion}

The name historian we describe in Section~\ref{sec:historicNameService}
achieves the primary goal of this work, reachability in the face of
identity mobility.  HINTS accomplishes this without the need for brand
new globally visible, unique, persistent identifiers for every mobile
person.  Furthermore, it requires no retrofitting of current name-space
providers' practices, requires no cooperation whatever from those
providers, and can be deployed incrementally.

Unfortunately, the simple centralized approach we take here has some
shortcomings that can become significant as an increasing fraction of
the world's business and fraud are transacted online.  First, it relies
on the honesty of the organization that operates the name historian.  A
centralized, centrally operated historian is a single point of failure,
for failures including corruption, malfunction, connectivity disruption
or going out of business.  Second, the scheme offers no tangible
assurances for the correctness of the information it gives out.  For
example, when a correspondent receives a primary name in response to a
resolution request, he has to take the answer on faith; he has to
believe that the historian did not invent the association and that the
historian checked the validity of the association (i.e., issued a
challenge to the claimed primary name and received a satisfactory
response).  Third, the scheme is only useful to those correspondents and
mobile people who trust the historian.  If no single historian is
globally trusted, then there is no straightforward way to allow any
correspondent to reach any mobile person.

Another issue that can potentially affect the utility of HINTS is its
reliance on the correspondents' knowledge of past names for a mobile
person.  If I never knew Jane Mobile before, there is no historic name
that I can use to resolve Jane's current contact information.  HINTS is
not intended as a search engine for finding contact information for
people I have never attempted to reach before.  We believe that web
search engines and currently deployed lookup services such as
WhoWhere~\cite{Whowhere} are more appropriate for that purpose.
However, old contact information gathered from such a web search becomes
more valuable with HINTS, since it can then potentially be resolved to
current contact information.

A third issue that affects the potential success of HINTS is concern for
the privacy of individuals.  Entering identifiers into a HINTS name
trail is voluntary; if a person wants obsolete identifiers to remain so,
they need not tell HINTS about those identifiers.  However, mobile
people need to be aware that there is a risk that messages sent by
correspondents to those old identifiers may reach someone else who now
has control over that old identifier.  If those messages are potentially
sensitive, then mobile people may desire to see a widespread use of
HINTS.  Furthermore, we believe that identifiers inevitably ``leak out''
to the outside world.  True protection from spam and other threats is
better performed using other techniques, such as the use of a personal
communications proxy through which all incoming communications may be
filtered~\cite{Roussopoulos1999}.

In the next section, we sketch an enhanced name historian design that
offers the same functionality as what we present in earlier sections,
but also alleviates the security and fault-tolerance concerns addressed
above.  We are currently focusing our design and implementation efforts
on this stronger version of HINTS.

\section{A Secure Name Historian}
\label{sec:secureHints}

We believe that the increasing need for online security, the rising cost
of identity theft and spoofing and the evolving standardization projects
for secure directory access will soon change how name-space providers do
naming.  We expect that providers will act much more like cooperative
but competing certification authorities in the future; in fact, most
large enterprises and even some governments~\cite{CanadaAddressChange}
do perform naming or digital certification within their own realms in
this more secure fashion.

Furthermore, we believe that unconditional trust in any centralized
service is going to be increasingly difficult to digest for
security-conscious mobile people.

In this section we enhance our earlier design for a historic name-trail
service to address two major concerns: First we use cryptography to
prevent illegitimate changes in name-to-person associations, as can be
caused by IP and email address spoofing, and eavesdropping on unsecured
email.  Second, we relax the need for trusting the name historian
unconditionally, by employing secure time stamping~\cite{Haber1991} and
undeniable attestation techniques~\cite{Buldas2002}, to limit the amount
of unobtrusive damage a corrupt historian can cause to its clients' name
histories.

We start by describing the stronger and slightly more cooperative
name-space providers we need for this enhanced design, and then outline
how our earlier concerns can be addressed.

\subsection{Certification Authorities As Name-Space Providers}
\label{sec:securehints:certificationAuthorities}

A certification authority is not much more than a name-space provider
that accompanies the assignment of a name to a person with the issuance
of a signed statement called an \emph{identity certificate}, such as the
one shown in Figure~\ref{fig:IdentityCertificate}.  Note that Jane
Mobile is not mentioned explicitly in the certificate.  Instead, the
provider assigns the name to the person who knows the secret portion of
the public key \key{AB34D9}, after having established that Jane Mobile
holds that key.  The kind of identity verification performed before a
name-space provider assigns a name to a person's public key is
provider-specific and out of scope in this paper.

\begin{figure}
\begin{verbatim}
IdentityCertificate{
  Issuer     : yahoo.com
  Subject    : jmobile
  Key        : AB34D9...
  Start Time : August 1, 1999,
  End Time   : July 31, 2001,
  Nonce      : 2C08A3...
  Signature  : <Issuer, subject, key,
                times and nonce signed
                using yahoo.com's
                private key>
}
\end{verbatim}
\caption[An identity certificate]{\small The identity certificate that
links the name \pn{jmobile}{yahoo.com} to the public key \key{AB34D9}.}
\label{fig:IdentityCertificate}
\end{figure}

The name-space provider can revoke an assignment by publishing a
\emph{revocation certificate}, such as that shown in
Figure~\ref{fig:RevocationCertificate}, or by publishing another
identity certificate for the same name but a different public key.

Finally, the name-space provider refreshes an assignment by issuing new
identity certificates for the same name and key before the previous
assignment expires.

\begin{figure}
\begin{verbatim}
RevocationCertificate{
  Issuer     : yahoo.com
  Subject    : jmobile
  Key        : AB34D9...
  Start Time : May 25, 2000,
  Nonce      : 69C802...
  Signature  : <Issuer, subject, key,
                start time and nonce
                signed using yahoo.com's
                private key>
}
\end{verbatim}
\caption[A revocation certificate]{\small The revocation certificate
that breaks the link between the name \pn{jmobile}{yahoo.com} and the
public key \key{AB34D9}.}
\label{fig:RevocationCertificate}
\end{figure}

Besides this basic certification functionality, in this environment of
higher security awareness we assume that name-space providers are
cooperative with the history service, in the sense that they notify the
historian of name assignment changes by conveying to the historian newly
issued identity or revocation certificates.  For replay protection, we
also assume that each such certificate issued by a name-space provider
contains a nonce value, which in the simplest case is the time of
issuance, but can also be a random number picked as a freshness
challenge.

\subsection{Certified Historic Naming}

One straightforward way to address this harder naming problem is to make
all information that the historian maintains signed, so as to prevent
illegitimate modifications.  To make sure that a corrupt historian
cannot divert a name trail by changing to what historian user accounts
point, mobile people are represented in name histories by their signing
key pairs, called \emph{person keys}, similarly to the approach taken in
SPKI/SDSI~\cite{RFC2693}.  In this manner, the online representation of
a person becomes ``the person who knows the secret signing key''.

Name assignments can be naturally represented with the identity and
revocation certificates that name-space providers issue.  Linking can be
represented with similar \emph{link} and \emph{severance certificates}
(see Figures~\ref{fig:LinkCertificate} and
\ref{fig:SeveranceCertificate}, respectively), signed by mobile people
(i.e., the holders of historian accounts) who lay claim on different
primary names.  Instead of storing a single association record per
primary-name-to-mobile-person association, the historian stores all
signed certificates delineating the assignment and the linking time
periods that make up an association.

\begin{figure}
\begin{verbatim}
LinkCertificate{
  Name       : jmobile@yahoo.com
  Person Key : E51BB2...
  Start Time : March 2, 2000,
  End Time   : May 1, 2000,
  Nonce      : 6FC3F0...
  Signature 1: <Name, Person Key,
                Times and Nonce
                signed by the current
                key of the name>
  Signature 2: <Name, Person Key,
                Times and Nonce
                signed by the current
                key of the historian
                account>
}
\end{verbatim}
\caption[A link certificate]{\small A link certificate associating name
\pn{jmobile}{yahoo.com} with the person currently represented by key
\key{E51BB2} on 3/2/2000.}
\label{fig:LinkCertificate}
\end{figure}

\begin{figure}
\begin{verbatim}
SeveranceCertificate{
  Name       : jmobile@yahoo.com
  Person Key : E51BB2...
  Time       : April 25, 2000,
  Nonce      : EE3BF4...
  Signature  : <Name, Person Key,
                Time and Nonce
                signed by the
                person key>
}
\end{verbatim}
\caption[A severance certificate]{\small A severance certificate
breaking the link between the person currently represented by the key
\key{E51BB2} and \pn{jmobile}{yahoo.com} on April 25, 2000.}
\label{fig:SeveranceCertificate}
\end{figure}

The historian can perform the same tasks as those in
Section~\ref{sec:historian} with only little more difficulty.  For
example, to resolve a historic name, the historian must find an
association from the historic name to a mobile person and then back from
the mobile person to a currently valid name association, as detailed in
Algorithm~\ref{alg:simpleResolution}.  Instead of just locating an
association record, the historian must find the appropriate certificates
justifying the association from the given historic name to a person and
back to another primary name; not only does the historian have to return
the answer it found---a primary name or a negative answer---but it must
also return a \emph{proof}, the set of certificates that support its
response.  The correspondent must then check that all the statements are
signed correctly, and that they in fact do support a valid resolution,
regardless of whether the answer was a primary name or a name-not-found
response.

Unfortunately, there are three major issues that face clients of such a
name historian: the short lifetime of digital signatures, the need for
temporal ordering of historic records, and the need for a ``closed'',
append-only historic database.  We elaborate on all three in turn.

First, digital signatures have a limited lifetime.  How does one make
sure that a statement signed a few years back was correctly signed at
the time, even though the signing key may have by now expired?  Some
work has been done to make signed documents usable even after the
pertinent signing key has expired~\cite{Haber1995,Maniatis2002}.  In any
case, the client must either trust the name-space providers themselves
to maintain historic records of when they used what public key pair to
sign their issued statements, or trust an external authority, such as
the KASTS Key Archival Service~\cite{Maniatis2002}, to maintain these
records.  Briefly, this service maintains a securely time stamped
archive of the keys that different name-space providers use during their
lifetimes, along with the times when those keys were used.  A client who
wishes to verify a signed statement, including the certificates we
describe above, can look up the name of the provider and the appropriate
time period.  The returned key can be then used to verify the signature
on the statement.  However, it is important to also ascertain that the
statement was signed \emph{while} the key found was still valid (i.e.,
before it expired or was revoked).

\begin{figure}
\begin{verbatim}
DelegationCertificate{
  Issuer     : E51BB2...
  Delegate   : D91452...
  Time       : June 1, 2001
  Nonce      : D8306A...
  Signature  : <Issuer, Delegate, Time
                and Nonce signed using
                both issuing and delegate
                keys>
}
\end{verbatim}
\caption[A delegation certificate]{\small A delegation certificate,
making the delegate key a continuation of the key trail of the issuing
key.}
\label{fig:DelegationCertificate}
\end{figure}

The ephemeral nature of digital signatures and signing keys also makes
it hard to maintain the online representation of mobile people when the
historian is not fully trusted, since mobile people must change the
public keys that represent them regularly.  People do this by issuing
and submitting \emph{delegation certificates} to the historian (see
Figure~\ref{fig:DelegationCertificate}, delegating their ``personhood''
from older keys to newer ones.  Again, timing is of the essence, since a
delegation must be performed before the previous key has to be
abandoned, due to expiration or compromise.

Second, any secure historic database must incorporate timing information
on when its different records were archived.  This is necessary for
archived signatures, as described in the previous paragraphs, but also
for authorization purposes, in key delegations or associations.  For
example, given a historic name, to establish the appropriate name
association from assignment and linking certificates one must have
knowledge of the relative temporal ordering of the issuance of those
certificates.  It must be shown that the historic name designates a time
\emph{after} an identity certificate and a linking statement created an
association between the pertinent name and mobile person, but
\emph{before} those certificates expire or any potential revocation or
unlinking statements came in effect.  Relative temporal authentication
of certificates in centralized~\cite{Buldas1998} and
decentralized~\cite{Maniatis2002b} certificate archives can be
invaluable here.

Briefly, relative temporal authentication uses one-way,
collision-resistant hash functions, such as SHA-1~\cite{SHA1}, to define
the temporal ordering from earlier historic records to later ones.  In a
sense, a tamper-evident linked list is created from all historic
records, so that earlier records appear earlier in the list.  Then
regularly picked placeholder records from the list are published in a
widely witnessed, secure, write-once publication medium, such as a
high-circulation newspaper; a verifier who cares about when a particular
record was appended into the historic database can trace the linked list
backwards and forwards to find the previous and next, respectively, list
links published in the newspaper, which in turn places the record in
question in a rough time frame, that is, between the publication dates
of two newspaper issues in the newspaper example.  The
collision-resistance and one-way property of the hash function used to
link records together guarantees that once a record has been placed in
the historic database and the next link from the database has been
``committed'' on a newspaper, neither the maintainer of the database nor
anyone else can tamper with the past, changing when records appear to
have been incorporated, adding or removing records, or modifying the
contents of those records.  This is the basic mechanism behind secure
time stamping~\cite{Haber1991}.

Third, it is important that the historian be unable to ``forget''
historic records that it has successfully accepted when they were
submitted to it.  For example, if a name-space provider submits a
revocation certificate to the historian and the historian accepts it, it
should be unable to deny the existence of such a revocation certificate
convincingly when queried later.  Undeniable
attestation~\cite{Buldas2002} is a cryptographic construct that allows
clients to verify the historian's claimed existence or non-existence of
certain records.  The basic idea there is to construct a sorted data
structure that allows \emph{undeniable attestations} on its contents,
that is, proofs that a particular element belongs or does not belong to
the data structure.  In the secure HINTS design, the database indices
are, in fact, undeniable attesters.

An essential requirement is that attestation proofs are significantly
shorter in size than the entire data structure itself.  Consider, for
example, a historic name database that holds a few billion certificates
for many names and many mobile people.  It would be extremely
unrealistic to have to look through every single record in such a large
database before being able to conclude no interesting revocation
certificate has been archived there.  The constructs from the work by
Buldas et al.~\cite{Buldas2002} and Maniatis and
Baker~\cite{Maniatis2002b} offer attestations with sizes logarithmic in
the number of records stored, which, for extremely optimistic lifetime
and popularity projections for a service like HINTS, never exceed
roughly 20 KBytes; this is quite an acceptable size for records one
expects to receive over the network once or twice a day.

\subsection{Status}

The reduced-trust version of the HINTS design, and a further completely
decentralized HINTS, are still under active development.  Although we
have already designed---or implemented from existed designs---the tools
and techniques we describe in the previous section, we have yet to
deploy and quantify the implementation of this more secure HINTS system.

\section{Related Work}
\label{sec:related}

The literature regarding naming in distributed systems is vast, so
in this section we confine
ourselves to a sampling of systems and products that attempt to provide
online names specifically for people.

An early design of a global name service~\cite{Lampson1986} describes
how a global hierarchical name space can be implemented across a
distributed set of computing resources.  Our approach is in a sense the
opposite: we assume there are reasons for many separate name spaces to
exist, and that people's online identifiers will move from one such name
space to another.  It is this identity mobility we address.

The new Internet domain names ending with \url{.name} are intended to
provide a flat global name space for individuals that is not associated
with particular employers and institutions.  These names could indeed
function as the primary names in our naming scheme.  However, there are
still many reasons why registrants for these names may end up changing
their online identifiers in any case.  People who fail to pay their
bills may lose access to their \url{.name} identifiers.  People who
change personal names, such as when getting married or taking stage
names, may want to reflect this by changing their online identifiers as
well.  Moreover, these online identifiers may not be the only online
identifiers registrants use.  A user of a \url{.name} identity may also
have email service through an employer, and it is hard to prevent
identifiers from these other name spaces from "leaking out" in such a
way that correspondents will not attempt to use them after they become
obsolete.

IdentiScape~\cite{Maniatis2002} describes a flat global name space for
people, in which identifiers do not necessarily reflect their personal
names but instead can be a set of space-separated words in unicode.  The
hope is that this name space is large enough to accommodate many names
per person, for everyone in the world, for the next one hundred years.
In the back end, IdentiScape maintains identity objects, which are
repositories of access-controlled personal information, such as
application-specific addresses, credit card numbers and user-specific
attributes.  In the model of Section~\ref{sec:model}, IdentiScape is a
two-tiered naming scheme.  The authors of IdentiScape list identifier
persistence as one of the desired traits of online identifiers, but
there is nothing to suggest that users will not change names within the
IdentiScape name space or use identifiers from other name spaces as
well, leaving identity mobility issues unsolved in IdentiScape.

\emph{CommonName}~\cite{CommonName} is one of many online redirection
services (others are OneName~\cite{OneName} and Novell's
DigitalMe~\cite{DigitalMe}).  CommonName allows the use of common names
or phrases instead of URLs or email addresses.  It also allows the owner
of a common name to set up different redirection schedules, for example,
the CommonName ``Jane Mobile'' is redirected to Jane's personal email
address during the weekend but to her work email address on a weekday.
Further personalization is possible, such as maintaining globally
accessible bookmarks, notes, etc.  CommonName provides plug-ins for
popular email applications and web browsers.  A common name is first
assigned according to availability, but then checked by a human for
``appropriateness''.  Common names must be ``appropriate and relevant to
[the] subscribers' web and e-mail resources.''

A CommonName is assigned to an account for the
duration of the account or the service:
\begin{quote}
CommonName is committed to providing this service free of charge for
as long as it is commercially viable for it to do so.
\end{quote}
However, once an account is discontinued, its CommonName can be
reassigned~\cite{Buchanan2002}.

\emph{PingID}~\cite{Pingid} is very similar to
IdentiScape~\cite{Maniatis2002}.  An identity server, privately held by
an owner of an identity, is responsible for authorizing (or not) the
release of information according to whomever is asking.  The business
goals cover the usual applications: single sign-on, password management,
privacy management.  PingID does not address identity mobility, in that
it does not allow reverse lookups from obsolete primary names to its own
name space.

\emph{Classmates.com}~\cite{Classmates} is a service that allows people
to register the school or college they attended, the military base at
which they served, or the company for which they worked.  This allows
people who, for example, attended the same class at the same school to
reconnect in the future.  HINTS has very similar goals to this venture.
We also seek to map a mobile person's presence in the past to a current
contact; however, we take a purely online and more general approach, in
that a correspondent need not know anything more than the mobile
person's identifier used in the past, rather than first and last names
and a class year.  This makes it easier for email and other applications
to run identifiers through HINTS automatically, perhaps every time they
are used, to make sure the most recent identifier is accessed, although
it limits the amount of heuristic weeding one can do on likely candidate
names.  In addition, HINTS provides a trail of past identifiers,
allowing correspondents throughout a recipient's history of identifiers
to contact the recipient, even if the correspondents did not know the
recipient when he graduated from Sample University in 1957.

\section{Conclusion}

The extremely volatile environment typical to online services and their
users makes identifier changes and the commensurate management problems
a painful fact of online life.  In this paper, we address the problem of
identity mobility, by extending the names people commonly use in today's
applications with a designation of a time when those names were
successfully used.

We present a simple design for HINTS, a historic name-trail service,
that allows mobile people to record and make available their movements
through the ``identifier landscape''.  Correspondents can follow those
name trails from where a person has been (i.e., a name she used to have)
to where that person is (i.e., a name she uses now).  In this way, a
correspondent can resolve a temporally qualified name into a name that
is valid now; he can use that name to reach a mobile person whom he has
not contacted in a long time, avoiding names that point to no one and
reassigned names that point to the wrong person.

Finally, we describe how security can enhance HINTS in the increasingly
naughty Internet to prevent illegitimate use of historic names and
name-history corruption by a malicious service itself.  We outline a
design for such an enhanced HINTS system, which we hope to evaluate and
make available soon.

{\footnotesize \bibliographystyle{acm} \bibliography{IdentiScape}}

\begin{thebibliography}{10}

\bibitem{Buchanan2002}
{\sc Buchanan, I.}
\newblock {CommonName Ltd.}
\newblock Personal Communication, May 2002.

\bibitem{Buldas2002}
{\sc Buldas, A., Laud, P., and Lipmaa, H.}
\newblock {Eliminating Counterevidence with Applications to Accountable
  Certificate Management}.
\newblock {\em Jounal of Computer Security 10}, 3 (2002), 273--296.

\bibitem{Buldas1998}
{\sc Buldas, A., Laud, P., Lipmaa, H., and Villemson, J.}
\newblock {Time-stamping with Binary Linking Schemes}.
\newblock In {\em Advances on Cryptology (CRYPTO 1998)\/} (Santa Barbara, USA,
  Aug. 1998), H.~Krawczyk, Ed., vol.~1462 of {\em Lecture Notes in Computer
  Science}, Springer, pp.~486--501.

\bibitem{CanadaAddressChange}
{\sc {Canada Customs and Revenue Agency}}.
\newblock {Address Changes Online}.
\newblock
  \url{http://www.ccra-adrc.gc.ca/eservices/tax/individuals/aco/menu-e.html},
  Sept. 2002.
\newblock An application of the Canadian \emph{Government On-Line} initiative,
  to provide every Canadian citizen with a digital ID.

\bibitem{Classmates}
{\sc {Classmates}}.
\newblock {The World's Best Place To Reunite}.
\newblock \url{http://www.classmates.com/}.

\bibitem{CommonName}
{\sc {CommonName Ltd.}}
\newblock {CommonName: Your internet identity}.
\newblock Available at \url{http://www.commonname.com/}.

\bibitem{RFC2693}
{\sc Ellison, C., Frantz, B., Lampson, B., Rivest, R., Thomas, B., and Ylonen,
  T.}
\newblock {RFC 2693}: {SPKI} certificate theory, Sept. 1999.

\bibitem{Haber1995}
{\sc Haber, S., Kaliski, B., and Stornetta, S.}
\newblock {How do Digital Time-stamps Support Digital Signatures?}
\newblock {\em CryptoBytes, {RSA} Laboratories 1}, 3 (Autumn 1995), 14--15.

\bibitem{Haber1991}
{\sc Haber, S., and Stornetta, W.~S.}
\newblock {How to Time-stamp a Digital Document}.
\newblock {\em Journal of Cryptology: the Journal of the International
  Association for Cryptologic Research 3}, 2 (1991), 99--111.

\bibitem{Lampson1986}
{\sc Lampson, B.~W.}
\newblock {Designing a Global Name Service}.
\newblock In {\em Proceedings of the Fifth Annual {ACM} Symposium on Princiles
  of Distributed Computing\/} (Calgary, {AL}, Canada, Aug. 1986), ACM,
  pp.~1--10.

\bibitem{Whowhere}
{\sc {Lycos, Inc.}}
\newblock {WhoWhere?}
\newblock \url{http://www.whowhere.lycos.com/}.

\bibitem{Maniatis2000}
{\sc Maniatis, P., and Baker, M.}
\newblock {IdentiScape: Tackling the Personal Online Identity Crisis}.
\newblock Technical Report CSL-TR-00-804, Computer Systems Laboratory, Stanford
  University, Stanford, {CA}, {USA}, June 2000.

\bibitem{Maniatis2002}
{\sc Maniatis, P., and Baker, M.}
\newblock {Enabling the Archival Storage of Signed Documents}.
\newblock In {\em Proceedings of the USENIX Conference on File and Storage
  Technologies (FAST 2002)\/} (Monterey, CA, USA, Jan. 2002), USENIX
  Association, pp.~31--45.

\bibitem{Maniatis2002b}
{\sc Maniatis, P., and Baker, M.}
\newblock {Secure History Preservation Through Timeline Entanglement}.
\newblock In {\em Proceedings of the 11th USENIX Security Symposium\/} (San
  Francisco, {CA}, {USA}, Aug. 2002), pp.~297--312.

\bibitem{RFC1034}
{\sc Mockapetris, P.~V.}
\newblock {RFC 1034}: Domain names --- concepts and facilities, Nov. 1987.

\bibitem{SHA1}
{\sc National Institute of Standards and Technology ({NIST})}.
\newblock {\em {Federal Information Processing Standard Publication 180-1:
  Secure Hash Standard}}.
\newblock Washington, {D.C.}, {USA}, Apr. 1995.

\bibitem{DigitalMe}
{\sc {Novell, Inc.}}
\newblock {DigitalMe}.
\newblock Available at \url{http://www.digitalme.com/}.

\bibitem{OneName}
{\sc {OneName Corporation}}.
\newblock {OneName}.
\newblock Available at \url{http://www.onename.com/}.

\bibitem{Pingid}
{\sc {PingID}}.
\newblock {Solving The Business Of Identity}.
\newblock \url{http://www.pingid.com/}.

\bibitem{RFC0821}
{\sc Postel, J.~B.}
\newblock {RFC 821}: Simple mail transfer protocol, Aug. 1982.

\bibitem{Roussopoulos1999}
{\sc Roussopoulos, M., Maniatis, P., Swierk, E., Lai, K., Appenzeller, G., and
  Baker, M.}
\newblock {Person-level Routing in the Mobile People Architecture}.
\newblock In {\em Proceedings of the 2nd {USENIX} {S}ymposium on {I}nternet
  {T}echnologies and {S}ystems\/} (Boulder, {CO}, {USA}, Oct. 1999), USENIX
  Association, pp.~165--176.

\bibitem{RFC2251}
{\sc Wahl, M., Howes, T., and Kille, S.}
\newblock {RFC 2251}: {Lightweight Directory Access Protocol} (v3), Dec. 1997.

\end{thebibliography}


\end{document}